\documentclass[showpacs, twocolumn,
 	amsmath, amssymb, aps, prl, 10pt]{revtex4-1}

\usepackage[utf8]{inputenc}
\usepackage{graphicx}
\usepackage{dcolumn}
\usepackage{bm}
\usepackage{siunitx}
\usepackage{hyperref}

\graphicspath{{figures/}}

\newcommand{\vE}{\bm{\mathcal{E}}}

\newcommand{\ket}[2][]{\mathinner{\lvert#2\rangle}_{\hspace{-0.1em}#1}}

\newcommand{\mean}[1]{\langle #1 \rangle}

\begin{document}

\title{Fictitious magnetic field gradients in optical microtraps as an experimental tool for interrogating and manipulating cold atoms}

\author{B.~Albrecht}
\author{Y.~Meng}
\author{C.~Clausen}
\author{A.~Dareau}
\author{P.~Schneeweiss}
\email{schneeweiss@ati.ac.at}
\author{A.~Rauschenbeutel}
\email{arno.rauschenbeutel@ati.ac.at}

\affiliation{%
 Vienna Center for Quantum Science and Technology,\\
 TU Wien -- Atominstitut, Stadionallee 2, 1020 Vienna, Austria
}%

\date{\today}

\begin{abstract}
Optical microtraps provide a strong spatial confinement for  laser-cooled atoms. They can, e.g., be realized with strongly focused trapping light beams or the optical near fields of nano-scale waveguides and photonic nanostructures. Atoms in such traps often experience strongly spatially varying AC Stark shifts which are proportional to the magnetic quantum number of the respective energy level. These inhomogeneous fictitious magnetic fields can cause a displacement of the trapping potential that depends on the Zeeman state. Hitherto, this effect was mainly perceived as detrimental. However, it also provides a means to probe and to manipulate the motional state of the atoms in the trap by driving transitions between Zeeman states. Furthermore, by applying additional real or fictitious magnetic fields, the state-dependence of the trapping potential can be controlled. Here, using laser-cooled atoms that are confined in a nanofiber-based optical dipole trap, we employ this control in order to tune the microwave coupling of motional quantum states. We record corresponding microwave spectra which allow us to infer the trap parameters as well as the temperature of the atoms. Finally, we reduce the mean number of motional quanta in one spatial dimension to $\mean{n}=\num{0.3(1)}$  by microwave sideband cooling. Our work shows that the inherent fictitious magnetic fields in optical microtraps expand the experimental toolbox for interrogating and manipulating cold atoms.
\end{abstract}

\maketitle

Optical dipole forces are a ubiquitous tool for trapping and manipulating ultracold atoms. Notable achievements with optical dipole traps include the investigation of quantum-degenerate gases~\cite{StamperKurn1998,Granade2002}, quantum simulation of many-body systems in optical lattices~\cite{Bloch2005}, long-lived quantum memories for light~\cite{Dudin2013}, and optical frequency standards and precision spectroscopy~\cite{Katori2003}. Optical microtraps confine a single or a few atoms to a small volume. They can, for example, be formed in the focus of a lens system with a high numerical aperture~\cite{Schlosser2001} and have recently been employed, e.g., to study Rydberg interactions~\cite{Wilk2010,Zhang2010} as well as the collisional entangling dynamics between two individual atoms~\cite{Kaufman2015}. Microtraps have also been created in the near field of optical nanofibers, i.e., cylindrical dielectric waveguides with a diameter that is smaller than the wavelength of the guided light. They offer a strong transverse confinement of the guided light over their full length and thereby enable a homogeneous and efficient coupling to ensembles of trapped atoms~\cite{Vetsch2010,Goban2012}. In addition, nanostructuring dielectric waveguides allows one to engineer their dispersion and to introduce photonic bandgaps in order to, e.g., drastically enhance the coupling between the atoms and the guided mode~\cite{Goban2014}. Nano-scale photonic waveguides  thereby open the route towards optical nonlinearities at the single photon level~\cite{Chang2014} as well as the study of new physical effects such as atom--photon bound states~\cite{Douglas2015}.

For atoms in optical traps, elliptical polarization components of the trapping light fields in general give rise to Zeeman state-dependent energy shifts that are equivalent to the effect of a fictitious magnetic field~\cite{CohenTannoudji1972}. A spatially varying intensity or ellipticity of the light then leads to a gradient of this fictitious magnetic field which has been employed to, e.g., demonstrate an optical analogue of the Stern-Gerlach effect~\cite{Chang2002}. Moreover, the intentional introduction of such fictitious magnetic field gradients into optical lattices enabled, e.g., the demonstration of coherent spin-dependent transport~\cite{Mandel2003}, quantum walks~\cite{Karski2009}, selective addressing of qubits~\cite{Lundblad2009}, and microwave sideband cooling in state-dependent potentials~\cite{Forster2009,Li2012}. In optical microtraps, a spatially varying elliptical polarization occurs naturally: According to Gauss' law applied to a propagating light field with a slowly varying envelope, the local ellipticity is a function of the divergence of the transverse field components~\cite{Bliokh2015,Lodahl2016}. Remarkably, when the transverse field components vary significantly on the wavelength scale, the local ellipticity can approach unity even when the transverse field itself is linearly polarized. The strongly inhomogeneous fictitious magnetic field that occurs in such situations is usually considered to be detrimental: It gives rise to a Zeeman state-dependent displacement of the trap minimum which may lead to dephasing, reduced fidelities in optical pumping schemes, and heating~\cite{Kaufman2012}. Therefore, so far, the effects of confinement-induced fictitious magnetic fields have typically been suppressed in experiments~\cite{Lacroute2012,Kaufman2012,Thompson2013}. 

Here, we take advantage of confinement-induced fictitious magnetic fields in order to couple the motional state of laser-cooled atoms in a nanofiber-based optical dipole trap with their internal hyperfine states via microwave transitions. We record corresponding microwave spectra which allow us to infer the trap parameters as well as the temperature of the atoms. Furthermore, by applying additional homogeneous real magnetic offset fields or fictitious magnetic field gradients created by an auxiliary light field, we control the state-dependent displacement of the trapping potential or even cancel the fictitious magnetic field, respectively. Based on these techniques, we tune the microwave coupling of motional quantum states. This allows us to establish favorable conditions for microwave sideband cooling which we then use in order to reduce the mean number of motional quanta along the direction of the displacement to $\mean{n}=\num{0.3(1)}$, i.e., \SI{77(6)}{\%} of the atoms reside in the motional ground state. Finally, by a time series of temperature measurements, we determine the heating rate of the trapped atoms. This shows that the inherent fictitious magnetic field in optical microtraps expands the experimental toolbox for interrogating and manipulating cold atoms.

Optical dipole traps rely on the intensity-dependent energy shift of the atomic ground state for an atom exposed to an optical field which is far-detuned with respect to the atomic ground to excited state transition frequencies. To calculate the energy shift, one usually decomposes the atomic polarizability into scalar ($\alpha_{\rm s}$), vector ($\alpha_{\rm v}$), and tensor components. For alkali atoms with ground-state angular momentum $J = 1/2$, the tensor polarizability vanishes, and we are left with the following interaction Hamiltonian~\cite{LeKien2013},
\begin{equation}
	\label{eq:light-shift}
	\hat{V} = -\frac{1}{4}\alpha_s |\vE|^2 + i\, \alpha_v
		\frac{\left(\vE^* \times \vE \right) \cdot \hat{\bm{F}}}{8F}.
\end{equation}
Here, $\vE$ is the positive-frequency electric field envelope and $\hat{\bm{F}}$ is the total angular momentum operator. Equation~\eqref{eq:light-shift} is valid when the interaction energy is small compared to the hyperfine splitting, such that $F$ is a good quantum number. The second term of Eq.~\eqref{eq:light-shift} is equivalent to a magnetic interaction $g_F \mu_B \bm{B}_\text{fict} \cdot \hat{\bm{F}}$ with a fictitious magnetic field,
\begin{equation}
	\label{eq:bfict}
	\bm{B}_\text{fict} = \frac{\alpha_v}{8 g_F \mu_B F} i \left(\vE^* \times \vE\right),
\end{equation}
where $g_F$ is the hyperfine Landé factor and $\mu_B$ the Bohr magneton. We note that the magnitude of the term $i(\vE^* \times \vE)$ is maximal for a circularly polarized field and zero for linear polarization.

\begin{figure*}
	\centering
	\includegraphics{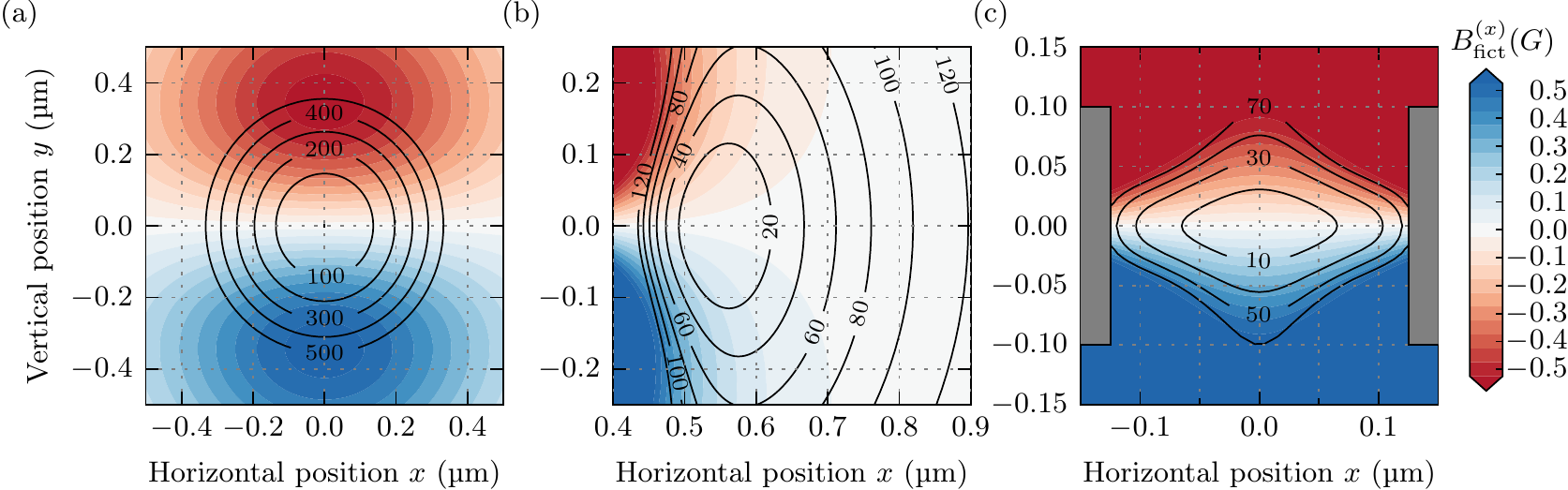}
	\caption{Examples of scalar trapping potential and fictitious field for three configurations of microtraps. The plots show transverse cuts through the trap minimum, where contours of the scalar potential are shown as black lines with labels in microkelvin, and the fictitious field is color-coded. For the latter, only the dominant $x$-component is shown. In all cases, the propagation direction of the relevant laser is into the plane (along $-z$), and the laser is polarized along $y$. (a)~Laser beam at a wavelength of \SI{850}{nm}, focused down to a waist radius of \SI{0.7}{\micro m} to trap $^{87}\text{Rb}$~\cite{Kaufman2012}. (b)~Two-color dipole trap for Cs atoms close to a nanofiber with \SI{250}{nm} radius whose axis is centered at $x=y=0$. Trap parameters are given in the main text. (c)~Trapping potential for Cs atoms generated by \SI{793}{nm} wavelength light propagating in the fundamental mode of a dual SiN nanobeam (gray) with \SI{250}{nm} gap, \SI{293}{nm} width and \SI{200}{nm} thickness. The parameters resemble those of Ref.~\cite{Hung2013}, but the photonic-crystal structure and Casimir-Polder potential have not been taken into account in the calculation.}
	\label{fig:bfict}
\end{figure*}

In general, fictitious magnetic fields cannot be neglected for propagating optical modes with a cross section comparable to the wavelength. This is exemplified in Fig.~\ref{fig:bfict} for three experimentally relevant cases of atoms in microtraps. Important insight can be gained for the simple case of a strongly focused Gaussian laser beam. Figure~\ref{fig:bfict}(a) shows a cross-section of the trapping potential for $^{87}$Rb atoms, generated by a laser field with a wavelength of $\SI{850}{nm}$, focused down to a waist of $\SI{0.7}{\micro m}$. These specific parameters are taken from~\cite{Kaufman2012}, but very similar configurations were also reported in~\cite{Sortais2007,Thompson2013}. The solution to Maxwell's equations can be written as a series in the diffraction angle $\theta=\lambda/(\pi w_0)$~\cite{Lax1975,Salamin2007}, where $w_0$ is the waist radius, and $\lambda$ the wavelength. The 0\textsuperscript{th}-order term $\vE^{(0)}$ is the well-known TEM solution to the paraxial equation with purely transverse polarization. Here, the latter is assumed to be parallel to the unit vector $\bm{u}_y$ along $y$. The 1\textsuperscript{st}-order correction is a longitudinal component $\vE^{(1)} \propto i y \theta \mathcal{E}^{(0)} \bm{u}_z$, which oscillates in quadrature to $\vE^{(0)}$. The interference of $\vE^{(0)}$ and $\vE^{(1)}$ thus gives rise to elliptical polarization that lies in the $y$-$z$-plane. In the focal plane ($z=0$) and for the TEM$_{00}$ mode, this elliptical polarization leads to a fictitious magnetic field that is perpendicular to the $y$-$z$-plane,
\begin{equation}
	\label{eq:bfict_gauss}
	\bm{B}_\text{fict}^{(1)} = \frac{\alpha_v}{8 g_F \mu_B F}
		\left(- 2 \bm{u}_x \,\theta\, \frac{y}{w_0} |\vE^{(0)}_{\rm max}|^2 e^{-2\frac{x^2+y^2}{w_0^2}} \right),
\end{equation}
where $\vE^{(0)}_{\rm max}$ is the value of $\vE^{(0)}$ at the origin. Using Eqs.~(\ref{eq:light-shift})--(\ref{eq:bfict_gauss}) we calculate the scalar part of the trapping potential as well as the fictitious magnetic field, both shown in Fig.~\ref{fig:bfict}(a). If we approximate the potential at $x=z=0$ by $ky^2/2 + gy$, where $k=\alpha_s |\vE^{(0)}_{\rm max}|^2/w_0^2$ and $g = g_F \mu_B m_F b_f$ with $b_f = dB_\text{fict}^{(1)}/dy|_{x=z=0}$, we find that the different Zeeman states have their respective potential minimum displaced by an amount $\Delta y = -g / k =  \frac{1}{4\pi}\frac{\alpha_v}{\alpha_s}\frac{m_F}{F}\lambda$. This must be compared to the vertical extent of the wavefunction $\sigma_y = [\hbar^2 / (k M)]^{1/4}$, where $M$ is the atomic mass. For the trap parameters used for Fig.~\ref{fig:bfict}(a), for example, $|\Delta y| \approx |m_F| \times \SI{12}{nm}$ and $\sigma_y \approx \SI{26}{nm}$. Hence, the intrinsic fictitious field has a significant impact on the trapping potential, which will be important for the manipulation of the internal and external states of the atom.

The situation is comparable for atoms trapped in the evanescent fields of dielectric waveguides. Although the specific geometry of the dielectric, and the even tighter field confinement, will give rise to additional components of the fictitious field, also here the dominant component is oriented perpendicular to the $y$-$z$-plane. For a two-color dipole trap for Cs atoms next to a nanofiber~\cite{Vetsch2010,LeKien2013a}, with parameters given below, the scalar potential and fictitious field are shown in Fig.~\ref{fig:bfict}(b). We note that the field here is solely due to the blue-detuned trap laser, as the standing wave formed by the red-detuned trap laser does not contribute.
As a final example, in Fig.~\ref{fig:bfict}(c) we show a trap for Cs atoms based on \SI{793}{nm}-wavelength guided light in a dual SiN nanobeam~\cite{Hung2013}, and we expect a comparable result for slotted nanofibers~\cite{Daly2016}.

In the scope of this work, we experimentally explore the case in Fig.~\ref{fig:bfict}(b), i.e., the nanofiber-based two-color dipole trap for Cesium atoms. For a detailed description of the experimental setup, see~\cite{Vetsch2010}. The nanofiber used in our experiment features a waist diameter of $\SI{500}{nm}$. The trapping potential is formed by a running-wave field with a free-space wavelength of \SI{783}{nm} and a power of \SI{17.1}{mW}, and a standing-wave field at \SI{1064}{nm} wavelength with a power of \SI{1.25}{mW} per beam. We use three pairs of coils to compensate stray magnetic fields, and to apply a homogeneous offset field along an arbitrary axis. By applying a microwave field at a frequency around the hyperfine splitting between the $F=4$ and $F'=3$ ground state manifolds, one can drive transitions from a state $\ket{F,m_F,n}$ to $\ket{F',m_F',n'}$. Here, $n$ and $n'$ are the respective quantum numbers for the motional states. The transition strength is given by the effective Rabi-frequency $\Omega_{n,n'}=\Omega\, C_{n,n'}$, where $\Omega$ is the bare Rabi frequency and $C_{n,n'}$ is the Franck-Condon factor~\cite{Forster2009}. For $n \neq n'$, one way to obtain nonzero $C_{n,n'}$ is to introduce a finite relative displacement of the potentials between the initial and final states. This leads to the appearance of sidebands in the microwave spectrum. For our system, the displacement is predominantly along $y$, and we expect to see sidebands corresponding to the trap frequency $\omega_y$ along this direction.

In order to obtain microwave spectra, we load atoms into the nanofiber-based trap in their $F=4$ hyperfine ground state. We then ramp up an external offset field to $B_0 = \SI{1.56}{G}$ with an angle $\phi = \SI{66}{\degree}$ with respect to the $x$-axis. After this, we find the majority of atoms in $m_F=-4$. A microwave pulse with a boxcar-shaped envelope at a given frequency transfers atoms in a specific Zeeman state from $F=4$ to $F'=3$. The pulse amplitude corresponds to $\Omega \approx 2\pi \times \SI{10}{kHz}$ and the pulse duration maximizes the population transfer on the carrier transition. Atoms remaining in $F=4$ are removed by applying a push-out laser beam. The fraction $A$ of transferred atoms is then estimated by optically pumping them back to $F=4$, followed by a measurement of the absorption of a weak nanofiber-guided light field on the cycling transition of the D2 line.

\begin{figure*}
	\includegraphics[]{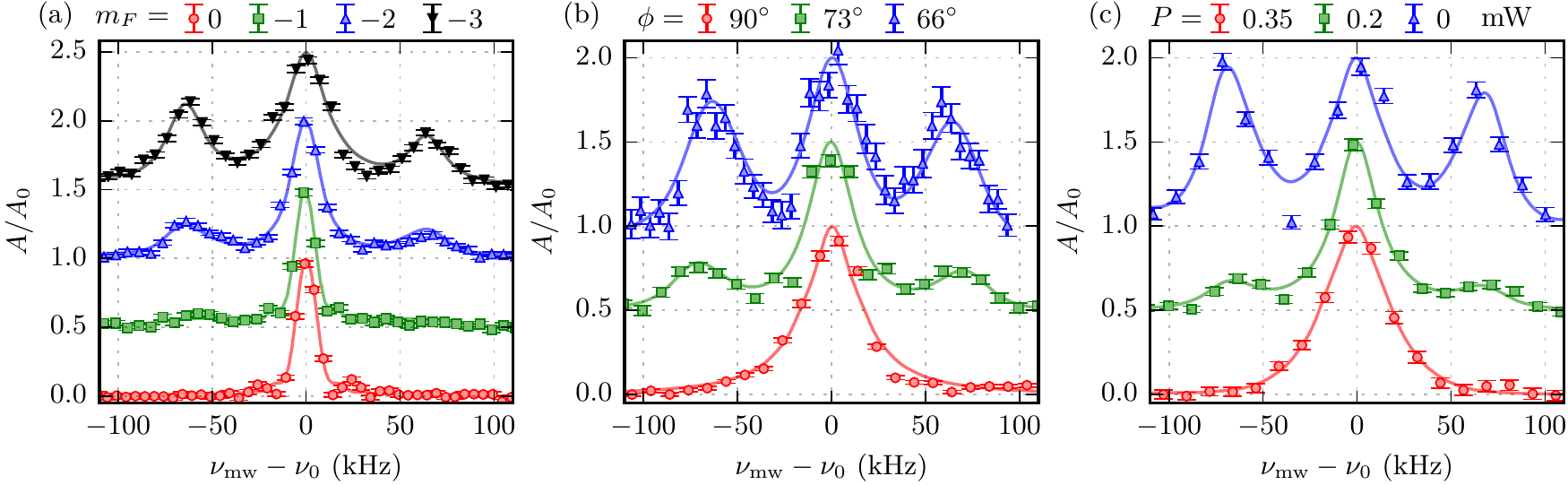}
	\caption{Microwave spectra for (a)~$\pi$-transitions between different Zeeman states, (b)~different angles of the offset magnetic field, and (c)~different powers of the tune-out laser. The ratio $A/A_0$ indicates the number of transferred atoms, normalized to the amplitude of the carrier. The microwave detuning is defined relative to the carrier transition, and different datasets are offset vertically for clarity. In (a) the initial state is $F=3$, while it is $F=4$ in (b) and (c). Solid lines are fits to an optical-Bloch-equation model (see text). Error bars indicate 1 standard deviation and stem from the non-linear fits used for atom-number estimation.}
	\label{fig:displacement}
\end{figure*}

As discussed above, the displacement $\Delta y$ of the trapping potentials depends on the Zeeman state and the gradient of the magnetic field. Figure~\ref{fig:displacement} shows a collection of microwave spectra that confirm these dependencies. Microwave spectra for $\pi$ transitions addressing different $m_F$ states are shown in Fig.~\ref{fig:displacement}(a). As an exception, here, the atoms are prepared in $F=3$ before taking the spectra. The zero of the detuning is defined as the center of the carrier transition, and the spectra are normalized to the amplitude $A_0$ of this transition. For $m_F=0$, we observe only a carrier transition. For $m_F\neq 0$, sidebands appear and become more pronounced for increasing $|m_F|$. The positions of the sidebands agree reasonably well with our calculated value $\omega_y = 2\pi \times \SI{77}{kHz}$, and we observe no sidebands at the two other trap frequencies $(\omega_x, \omega_z) = 2\pi \times (128, 198)\,\SI{}{kHz}$. 

Figures~\ref{fig:displacement}(b) and~(c) show two means to tune the strength of the sidebands for a given transition. For a sufficiently large offset field, $\Delta y$ is proportional to the derivative of the magnitude of the total magnetic field $\bm{B}_\text{tot} = \bm{B}_\text{fict} + \bm{B}_\text{off}$. An offset field at angle $\phi$ hence reduces the displacement by a factor $\cos \phi$. This is demonstrated in Fig.~\ref{fig:displacement}(b), which shows the transition from large to vanishing sideband amplitude when $\phi$ is varied between \SIlist{66;90}{\degree}. For every setting of $\phi$, we slightly adjusted the polarizations of the trap lasers in order to minimize the width of the microwave transitions. Another possibility to tune the displacement is by changing $\bm{B}_\text{fict}$ itself, using a nanofiber-guided laser operating at the tune-out wavelength of about \SI{880}{nm}~\cite{Arora2011}. At this wavelength, the scalar polarizability vanishes, and the scalar potential is not affected. The vector polarizability has the opposite sign than for light at \SI{783}{nm}, such that the total fictitious field is reduced when the two fields are co-propagating and identically polarized. One can clearly see from Fig.~\ref{fig:displacement}(c) the dependence of the sideband height on the applied tune-out laser power. For \SI{0.35}{mW}, the fictitious field at the position of the atoms is almost completely compensated, which fits very well with the theoretical expectation of about \SI{0.36}{mW}.

\begin{figure*}
	\makebox(0,0)[bl]{\includegraphics[]{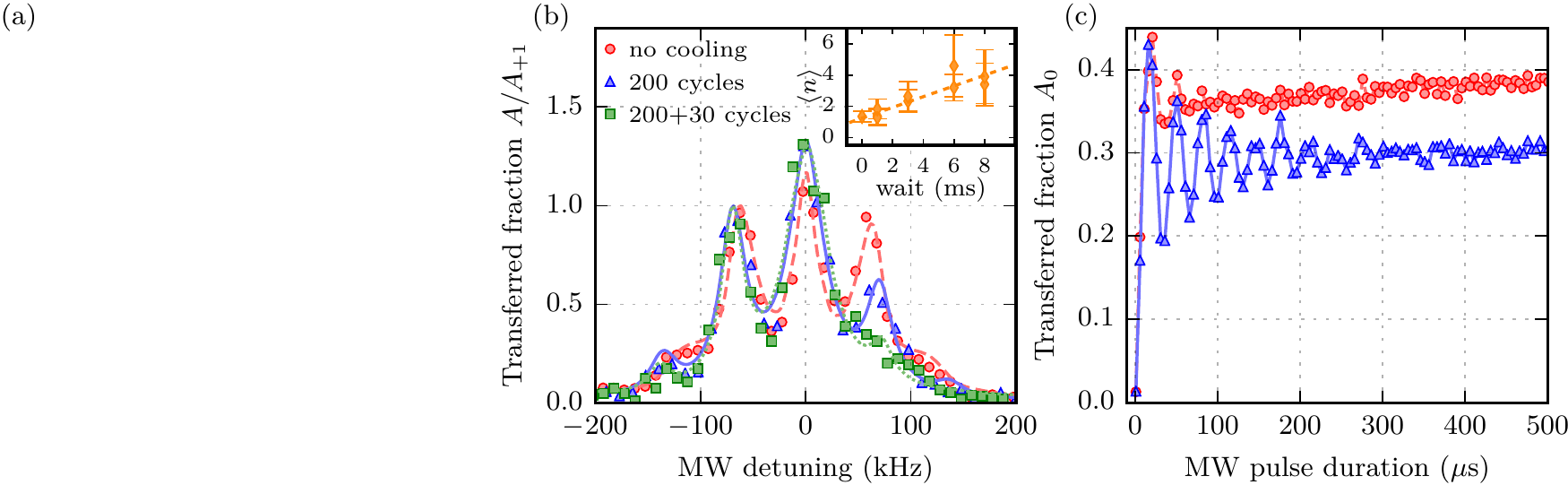}}
	\hspace{-1em}\includegraphics[]{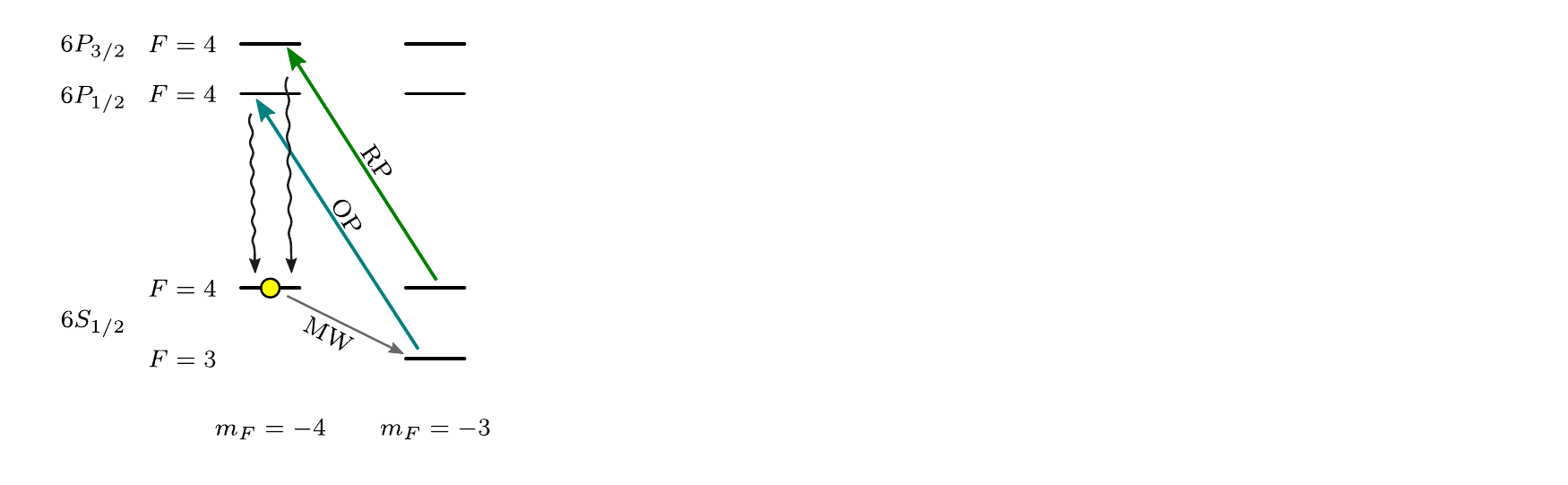}
	\caption{Microwave sideband cooling scheme and results. (a)~Diagram of our cooling scheme with relevant levels and transitions for Cesium. One cooling cycle consists of a microwave (MW) pulse and optical pumping (OP, RP) followed by spontaneous emission. (b)~Experimental microwave spectra on the $\ket{F=4,m_F=-4} \to \ket{F=3,m_F=-3}$ transition before cooling (red circles), after the application of 200 cooling cycles (blue triangles), and for 30 additional cycles without RP (green squares). The spectra are re-centered to compensate for a temperature-dependent shift of the transition frequency, on the order of $\SI{20}{kHz}$, that can be attributed to an $x$-dependent differential light shift induced by the trapping lasers. Spectra are normalized to the first left sideband. Lines are fit results (see text). The inset in (b)~shows the evolution of the mean number of motional excitations for a variable waiting time following the cooling sequence. A linear fit (dashed line) gives a heating rate of $\SI{0.34(1)}{quanta\per\milli\second}$. (c)~Rabi oscillations on the carrier of the $\ket{F=4,m_F=-4} \to \ket{F=3,m_F=-3}$ transition before (red circles) and after (blue triangles) 200 cooling cycles. Error bars in (b) and (c) are smaller than the symbol size.}
	\label{fig:cooling}
\end{figure*}

The temperature of the atomic ensemble can be estimated via the ratio of the amplitude of the sidebands. Assuming an ideal harmonic oscillator, the mean number of motional excitations is given by $\mean{n} =$ \mbox{$A_{-1}/(A_{+1}-A_{-1})$}, where $A_{m}$ is the amplitude of the sideband for the transition $n \to n + m$. In our case, the relevant sidebands have almost equal amplitude, meaning that $\mean{n} \gg 1$. However, the fact that we can resolve the sidebands in the microwave spectra enables us to implement cooling. With the atoms initially in $\ket{F=4, m_F=-4}$, a single cooling cycle, sketched in Fig.~\ref{fig:cooling}(a), consists of a \SI{20}{\micro s}-long microwave pulse on the $n\to n-1$ sideband with bare Rabi frequency around $2\pi\times\SI{40}{kHz}$. The atoms are then optically pumped back to the initial state using a $\sigma^-$-polarized light-field on the Cs D1 line, labeled OP. During the optical pumping, the atoms are brought back to the initial state, but also have a finite probability to spontaneously decay into $\ket{F=4, m_F=-3}$. Those atoms are re-integrated into the cooling cycle with a $\sigma^-$-polarized repumping field, labeled RP, on the Cs D2 transition. Both OP and RP are on for \SI{10}{\micro s}, and we leave OP on for another \SI{10}{\micro s} to pump all atoms out of $F=3$.

Figure~\ref{fig:cooling}(b) shows microwave spectra before and after 200 cooling cycles. The spectra are normalized to the amplitude $A_{+1}$ of the $n\to n+1$ sideband. As expected, the cooling results in a relative reduction of $A_{-1}$. To obtain quantitative information about the cooling efficiency, we fit the experimental data using an optical Bloch equation model. The model assumes a 1-dimensional potential and a thermal distribution of motional excitations. It takes into account the calculated anharmonicity of the trap and a finite dephasing time for the microwave transitions. Free parameters for the fit are the mean excitation number $\mean{n}$, trap frequency $\omega_y$, bare Rabi frequency $\Omega$, displacement $\delta=\Delta y' - \Delta y$ between the initial and final potentials, a finite dephasing time $T_2$, as well as an overall amplitude and a horizontal offset. As can be seen in Fig.~\ref{fig:cooling}(b), the fit reproduces well the shape of the spectra both without and with cooling. The obtained values for $\omega_y$, $\Omega$ and $T_2$ agree well with expectations, and within the errors they are the same for both data sets. For the data with cooling, we get $\delta/\sigma_y = \num{0.56(6)}$, in perfect agreement with the calculated value of $0.56$. Without cooling, the fit gives $\delta/\sigma_y = \num{0.34(2)}$. We attribute this discrepancy to the fact that hotter atoms are on average further away from the fiber~\cite{Vetsch2010}, where the scaling of the effective scalar potential and fictitious field along $y$ leads to a reduced displacement. Finally, the fit allows us to extract a mean excitation number of $\mean{n}=\num{10(2)}$ before, and $\mean{n}=\num{1.4(3)}$ after cooling, corresponding to a temperature around \SI{6}{\micro\kelvin}. In order to understand why we do not reach lower temperatures, we measured the background heating rate in our system by introducing a variable waiting time after the last cooling cycle, see inset of Fig.~\ref{fig:cooling}(b). We extract a heating rate of $\SI{0.34(1)}{quanta \per ms}$ from a linear fit. This heating rate should not be a significant limitation for the minimum achievable temperature. Instead, we identified excessive photon scattering as the main limitation: The OP and RP beams are mutually parallel and, for technical reasons, propagate at an angle of \SI{20}{\degree} to the offset magnetic field, which sets an upper limit on the obtainable degree of $\sigma^-$-polarization. As a consequence, the state $\ket{F=4, m_F=4}$ is not the desired dark state of the cooling sequence. We confirm this hypothesis by adding 30 cooling cycles where the RP light is off, see green data points and associated fit in Fig.~\ref{fig:cooling}(b). We then obtain a mean excitation number of $\mean{n} = \num{0.3(1)}$, at the expense of losing atoms to other $m_F$ states.

As another indication of successful cooling, we record Rabi oscillations between the states $\ket{F\!=\!3, m_F\!=\!-3}$ and $\ket{F\!=\!4, m_F\!=\!-4}$ by applying a microwave pulse of variable duration on the carrier transition before and after 200 cooling cycles. Figure~\ref{fig:cooling}(c) shows the transferred population as a function of pulse duration. Without cooling, the visibility of the Rabi oscillations decreases rapidly because all the motional states involved have a different effective Rabi frequency. After cooling, the mean phonon number is lowered and only few motional states contribute. As a consequence, the dephasing time is increased by an order of magnitude.

In summary, we investigated the strongly inhomogenous fictitious magnetic fields that arise from the tight confinement of light fields used to create optical microtraps. Taking advantage of these fictitious fields, we demonstrated techniques that allow one to probe and to manipulate atoms. Using microwave spectroscopy, we showed for the specific case of nanofiber-trapped Cesium atoms that the resulting state-dependent potentials can be tailored with external magnetic fields or with additional fiber-guided light. We exploited the state-dependent potentials to probe important parameters of the trap and the atoms therein. Performing microwave sideband cooling, we approached temperatures close to the motional ground state. Besides its general relevance for experiments with atoms in optical microtraps, the cooling technique may in particular facilitate studies of self-organization~\cite{Chang2013, Griesser2013} and lateral light forces~\cite{Scheel2015, RodriguezFortuno2015, Sukhov2015, Kalhor2016} for atoms close to waveguides. In addition, it may provide a well-defined starting point for loading atoms into surface-induced potentials~\cite{Chang2014a} or for the investigation of collapse and revival dynamics in nanofiber-based traps~\cite{LeKien2013b}. Furthermore, our work gives a first example of the utility of the inherent fictitious magnetic fields in optical microtraps. We believe that our approach can be extended to oscillating fictitious fields, enabling purely optical implementations of, e.g., microwave transitions and RF-dressed potentials. Such techniques may be advantageous for addressing and manipulating atoms near photonic nanostructures.

\begin{acknowledgments}
We thank Igor Mazets for helpful discussions, and Clement Sayrin for valuable insights, discussions and feedback during the preparation of the experiment. Financial support by the Austrian Science Fund (FWF, SFB NextLite Project No. F 4908-N23 and DK CoQuS project No. W 1210-N16) and the European Research Council (Consolidator Grant NanoQuaNt) is gratefully acknowledged. C.~C. acknowledges support by the European Commission (Marie Curie IF Grant No. 658556).
\end{acknowledgments}

\bibliography{sidebands}

\end{document}